\documentclass[aps,showpacs,prl,superscriptaddress,twocolumn]{revtex4}  

\usepackage{amsfonts}
\usepackage{amsmath}
\usepackage{mathrsfs}
\usepackage{amssymb}
\usepackage{wasysym}
\usepackage{subfigure}
\usepackage{float}
\usepackage{graphicx}
\usepackage{dsfont}
\usepackage{cases}
\usepackage{rotating}         
\usepackage{color}

\begin{document}

\title{Logistic growth of skyrmions in 3D chiral magnets}
\author{Jinl\"u Cao}
\affiliation{Department of Physics, Shanghai University, Shanghai 200444, P.R. China}
\affiliation{Qian Weichang College, Shanghai University, Shanghai 200444, P.R. China}
\author{Ying Jiang}
\thanks{Corresponding author}
\email{yjiang@shu.edu.cn}
\affiliation{Department of Physics, Shanghai University, Shanghai 200444, P.R. China}
\affiliation{Qian Weichang College, Shanghai University, Shanghai 200444, P.R. China}
\affiliation{Shanghai Key Laboratory of High Temperature Superconductors, Shanghai, China}
\affiliation{Key Lab for Astrophysics, Shanghai 200234, P.R. China}

\begin{abstract}
    The process of topological phase transition in a chiral magnetic material from skyrmion lattice phase to helical phase has been investigated numerically and experimentally, yet the analytical expression of the evolution of skyrmion number and emergent magnetic monopole charge during the process is still waiting to be explored. In this letter, by utilizing the topological current theory, we show that the change of skyrmion number of a layer in the system is equal to the net monopole charges that flows through the layer. Based on this relation, with the help of statistical argument, we derive the analytical expressions of the skyrmion number and the monopole charge as  functions of the external magnetic field. We find that the evolution of the skyrmion number is exactly the logistic growth function, and the evolution of the monopole charge is proportional to the derivative of the skyrmion number of the system. our analytical results are in good agreement with the numerical simulations in \cite{2013SCIENCE}. The goodness of data fitting goes up to 99.94\%.
\end{abstract}


\maketitle

Due to potential applications in spintronics and quantum information \cite{2013Writing and Deleting Skyrmions,2015_skyrmion_logic_gates,2015Magnetic bubbles with a twist}, skyrmions have been attracting numerous efforts in recent years \cite{2006skyrmion ground states, 2014Topological_Changes, 2015Capturing skyrmion, 2015Blowing skyrmion bubbles,2014dynamics of driven skyrmion, 2015 switching skyrmion, 2015 skyrmion size and shape,2015skyrmion dynamics,2013NNano}.

Skyrmions are emergent topologically protected structures \cite{1983_Wilczek_Zee,1988_Dzyaloshinskii_Polyakov_Wiegmann,1991Dynamics of Magnetic Vortices} which can naturally arise in magnetic thin films \cite{2010Lorentz microscopy,2011spin-resolved STM,2014_taylor_skyrmion_2D_Nat_Com} or in bulk magnets \cite{2009skyrmion lattice,2014_skyrmion_crystal_stiffness,2015_observation_skyrmion_lattice} when the
temperature of the system is sufficiently below the Curie point. In 2D thin films, a skyrmion is whirl-like magnetic configuration
consisting of a core and its surrounding magnetization vector field, and its core is a small region with magnetization inside pointing
perpendicular to the plane of the thin film. A 3D bulk magnet can be looked upon as a stack of magnetic thin film layers, in a small
external magnetic field, skyrmions in adjacent layers will align on top of each other, forming skyrmion lines oriented parallel to the
small external magnetic field.

In order to describe the topological properties of a magnetic configuration in a magnetic system, a skyrmion number \cite{1983_Wilczek_Zee,1988_Dzyaloshinskii_Polyakov_Wiegmann,1991Dynamics of Magnetic Vortices} for a 2D magnetic
thin film or a layer perpendicular to the external magnetic field in 3D bulk magnet is defined as follows
\begin{equation}
 Q^{\rm Skyrmion}  = \frac{1}{4 \pi} \int \vec{m}\cdot\left( \partial_{x} \vec{m}\times \partial_{y} \vec{m}\right) dxdy
 \label{skyrmion_number0}
\end{equation}
with $\vec{m}=\vec{M}/\|\vec{M}\|$ being the normalized magnetization vector field and $x$ and $y$ being the spatial coordinates of the
2D magnetic thin film or the layer in the 3D magnet. In fact, this is exactly a topological number called Pontryagin index, it is a integer number
and a topological invariant for a continuous manifold, in other words, the skyrmion number will not be changed under continuous deformation of the unit vector $\vec{m}$.

However, since the real magnetic material possess lattice structure, microscopically, $\vec{m}$ can only be well-defined on lattice sites, possible singularities off lattice-sites may be involved, and the skyrmion number will be changed mediated by emergent monopoles \cite{1988_Haldane,2009YJ}. This has been proven by recent experiment \cite{2013SCIENCE}. After preparing
the skyrmion lattice in a chiral magnet in external magnetic field, it is observed that, by changing the external field, two skyrmions may merge into one or single
skyrmion may split into two, during these processes, singular magnetic configurations locate exactly at the points of coalescence and separation. These singular configurations
are indeed the so-called emergent magnetic monopoles or antimonopoles \cite{2016Nature Comm}. Thus, by changing the external magnetic field, topological phase transition associated with the change of
skyrmion number has been achieved. Hence the relations between skyrmion number as well as the topological charge of monopoles and external magnetic field deserve to be investigated in detail. Actually, numerical simulation on this topic has been conducted \cite{2013SCIENCE}, yet the analytical expression of the curves of skyrmion number and monopole
charge along with the changing of external magnetic field has not been unveiled, and this is the main goal of the present letter.

In a 2D magnetic thin film, the skyrmion number in Eq.(\ref{skyrmion_number0}) can be written in a more symmetric way as
\begin{align}
 Q^{\rm Skyrmion} 
 \label{skyrmion_winding_number1} = &\frac{1}{8 \pi}
 \int \epsilon ^{0\mu\nu}\epsilon_{abc} m^a \partial_{\mu} m^b \partial_{\nu} m^c dxdy
\end{align}
with the Greek letters $\mu$ and $\nu$ standing for the two-dimensional spatial indices and the Latin letters $a$, $b$ and $c$ for the spin indices, the index $0$ denotes the time parameter $t$.
The topological charge density of skyrmions can further be re-expressed in terms of magnetization vector field $\vec{M}$ as \cite{2004YJ,2009YJ}
\begin{equation}
q=\frac{1}{8 \pi}
 \epsilon ^{0\mu\nu}\epsilon_{abc} \frac{M^a}{\|M\|} \partial_{\mu} M^b \partial_{\nu} M^c.
\end{equation}
By the use of the topological current theory \cite{2000YJ,2004YJ} and the Laplacian Green's function relation, straightforward calculation shows that
\begin{equation}
\partial_t q=D\left(\frac{M}{z}\right)\delta(\vec{M}),
\label{changerate-skyrmion-density}
\end{equation}
where $D(M/z)$ is the Jacobian of $\vec{M}$ and $\vec{z}=(t,x,y)$. Eq. (\ref{changerate-skyrmion-density}) clearly shows that the change rate of the skyrmion number is zero when $\vec{m}$ is well-defined in the whole space-time, i.e. when $\vec{M}$ possesses no zero point. However, the existence of singularities of $\vec{m}$ does change the skyrmion number. These singularities are the so-called emergent monopole events which are located at the zero points $\vec{z}_i=(t_i,x_i,y_i)$ of $\vec{M}$, i.e. $\vec{M}(\vec{z}_i)\equiv0$. Hence it is not difficult to recognize that after a time interval $\Delta t$ we have
\begin{equation}
\Delta Q^{\rm Skyrmion}=\sum_i Q^{\rm Monopole}_i,
\end{equation}
i.e. the change of the skyrmion number in this time interval is equal to sum of the topological charges of all monopoles appearing during the time interval. In fact, the topological charge $Q_i$ of the $i$-th monopole is exactly the wrapping number $W_i$ of unit vector field $\vec{m}$ at its $i$-th singular point $\vec{z}_i$ \cite{Hatcher_topology}, due to the second homotopy of sphere $\pi_2(S^2)=Z$, it is an integer.

For a 3D chiral magnet in external magnetic field, the above relation also holds true. Suppose the external field is along the $z$ direction, in a layer perpendicular to the external field, the change of the skyrmion number after a time interval reads
\begin{equation}
\Delta Q^{\rm Skyrmion}(z)=\sum_i Q^{\rm Monopole}_i(z),
\label{relation-charge-skyrmion-monopole}
\end{equation}
the right-hand-side of the equation indicates the sum of the charges of monopoles going through the layer from below during this time interval. It should be pointed out that an extra minus sign should be inserted in front of the monopole charge if the monopole goes through the layer from above.

Based on the above relation, we are going to investigate the behavior of the skyrmion number as well as the monopole number in the system from a statistical perspective. To do so, some definitions and statistical argument should be made first.

1. Due to the homogeneity of the system in the bulk, an average skyrmion number $Q^S(t)$ per layer at time $t$ can be defined as
\begin{equation}
Q^S(t)=\frac1L\int_0^L Q^{\rm skyrmion}(t,z)dz,
\end{equation}
$L$ is the thickness of the bulk material. We denote the maximum number of skyrmions allowed in the system per layer by $K$, and the average initial skyrmion number per layer is $Q_0$

2. Since the monopoles and antimonopoles are associated with splitting and annihilation of skyrmions, hence statistically speaking, monopoles (MP) and antimonopoles (AMP) should be homogenously distributed in the bulk of the system. Accordingly, we denote the monopole number and antimonopole number per unit thickness by $Q^{MP}$ and $Q^{AMP}$, respectively. In fact, the emergent monopoles (antimonopoles) experience forces coming from the external magnetic field \cite{2016Nature Comm,2014Dynamics of Emergent Magnetic Monopole}, causing MPs and AMPs flows along the $z$-direction oppositely, and since the MPs and AMPs possess opposite charges, their topological currents are along the same direction. Due to the spatial inverse symmetry of the system, it is safe to take the their average drift velocities to be $v_{MP}=-v_{AMP}\equiv v_{\rm drift}$. Consequently, we see that during the time interval $\Delta t$, the total charge of the monopoles flow through a layer is equal to $(Q^{MP}+|Q^{AMP}|)v_{\rm drift}\Delta t$. Since the relation in Eq. (\ref{relation-charge-skyrmion-monopole}) is also true when taking average with respect to $z$ on both sides, we then have
\begin{equation}
\Delta Q^S = (Q^{MP}+|Q^{AMP}|)v_{\rm drift}\Delta t.
\label{average-relation}
\end{equation}

3. For a large system, it is reasonable to assume that every two skyrmions  are equally likely to merge and create a MP/AMP in its merging point, thus the number of MP/AMP in the bulk is proportional to the number of existing skyrmions
\begin{equation}\label{Qm_proportional 1}
 Q^{MP/AMP} \propto  Q^{S}(t).
\end{equation}
Meanwhile, since in the system every annihilated skyrmion is associated with a MP/AMP, hence the number of MP/AMP is also proportional to the number of skyrmions which have already been annihilated, i.e.
\begin{equation}\label{Qm_proportional 2}
   Q^{MP/AMP} \propto  Q^{S}(t)^ {\{\rm annihilated\}}= K -  Q^{S}(t).
\end{equation}
By combining Eqs. \eqref{Qm_proportional 1} and \eqref{Qm_proportional 2}, and setting the coefficients to be $r_1$ and $r_2$ for MP and AMP respectively, we have
\begin{equation}\label{MP propto Qbar}
 |Q^{MP/AMP}| = r_{1,2}   Q^{S}(t)(K -  Q^{S}(t))
\end{equation}

With all the necessary information in hand, we can now go further to get the evolution functions of the skyrmion number and MP/AMP numbers.

Substitute Eq. \eqref{MP propto Qbar} into Eq. \eqref{average-relation}, and then take the limit of $\Delta t\rightarrow 0$, we get
\begin{equation}\label{ODE Q Skyrmion 2}
\frac{dQ^S}{dt}=r Q^{S}(t)\left(1 - \frac{ Q^{S}(t)} {K}\right),
\end{equation}
where $r =(r_1 + r_2) v_{\text{drift}} K$.
The solution of this differential equation is
\begin{equation}\label{Qbar curve}
   Q^S(t) = \frac{K  Q_0 e^{rt}}{K +  Q_0(e^{rt} -1)}.
\end{equation}
We find that the evolution function of skyrmion number of the 3D chiral magnetic system is exactly the logistic function!

Straightforwardly, we have for the charge of MPs/AMPs
\begin{equation}\label{ODE Q monopole 2}
    Q^{MP/AMP}
    = \pm     \frac{ r_{1,2 }K}{r} \frac{d  Q^{S}(t)}{d t},
\end{equation}
the plus and minus signs in the right-hand-side correspond to MP and AMP, respectively.
Together with Eq. \eqref{Qbar curve}, this leads to the expression of MP/AMP charges in the bulk,
\begin{equation}\label{Q monopole 1}
Q^{MP/AMP} = \pm\frac{ r_{1,2 }K}{r}
                \frac{r K  Q_0(K- Q_0)  e^{-rt}}{ [(K- Q_0)  e^{-rt} +  Q_0]^2}.
\end{equation}

It should be pointed out that the system will stay in an equilibrium state when keeping the external physical condition unchanged. In order to observe the evolution of the skyrmion number as well as monopole numbers in experiment and in numerical simulation \cite{2013SCIENCE}, the external magnetic field is set to vary slowly, the process is sufficiently slow to keep the metastable nature of the system.

In order to obtain the relations between the topological charges and the external magnetic field, the time parameter $t$ has to be replaced by $B$. Since the magnetic field changes with a very slow constant rate, i.e.
$$ \frac{dB}{dt} = \frac{1}{h} = \text{const.}
$$
 Therefore, the topological charges, expressed in terms of the field strength $B$, are
 \begin{equation}\label{Qbar curve2}
   Q^S(B) = \frac{K  Q_0 e^{rhB}}{K +  Q_0(e^{rhB} -1)},
\end{equation}
and
\begin{equation}\label{Q monopole 2}
  Q^{MP/AMP} (B) = \pm \frac{r_{1,2} K^2}{4} \text{sech}^2\left[\frac{rh}{2}(B-B_0)\right],
\end{equation}
where $B_0 = \frac{1}{rh} \text{ln}(\frac{K}{Q_0}-1)$.

We should keep in mind that Eq.\eqref{Qbar curve2} only describes the topological charge of the skrymions in the bulk.
In order to make a direct comparison with the numerical simulation for the skyrmion number on the surface of the material \cite{2013SCIENCE}, let us go back to Eqs. \eqref{average-relation} and \eqref{MP propto Qbar}. We see that
Eq. \eqref{MP propto Qbar} is not valid for both MPs and AMPs simultaneously at the upper or lower boundaries of the bulk. As is known, due to the drift caused by the external magnetic field, AMPs near the lower boundary move towards the outside of the bulk, then lose their topological protection and vanish. Therefore, on the lower surface, Eq. \eqref{MP propto Qbar} is only valid for MPs, while the AMPs are nearly absent, i.e.
\begin{equation}\label{MP propto Q on surface}
\left\{
\begin{aligned}
& Q^{MP} = r_{1} Q^{S}(t)(K -  Q^{S}(t)),\\
& Q^{AMP} \approx 0,
\end{aligned}
\right.
\;\;{\rm at}\; z = 0.
\end{equation}
The same situation is faced on the upper surface where the topological charge of MPs is nearly zero. By taking all this into account, we finally obtain that on the surface of the system the skyrmion number is expressed in terms of $B$ as
\begin{equation}
Q^S_{\rm boundary}(B)  = \frac{K' Q'_0 e^{\tilde rhB}}{K' + Q'_0(e^{\tilde rhB} -1)}
            + Q^s_{\rm correction},
\label{surface-skyrmion-number}
\end{equation}
where $Q^s_{\rm correction}={\rm const}.$ is a correction term, $K'$ and $Q_0'$ as well as $\tilde{r}$ are revised parameters. In other words, the evolution of skyrmion number on both surfaces satisfies the modified logistic growth curve in Eq. \eqref{surface-skyrmion-number}.

\begin{figure}[h!]
  \centering
  \includegraphics[width=0.72 \linewidth]{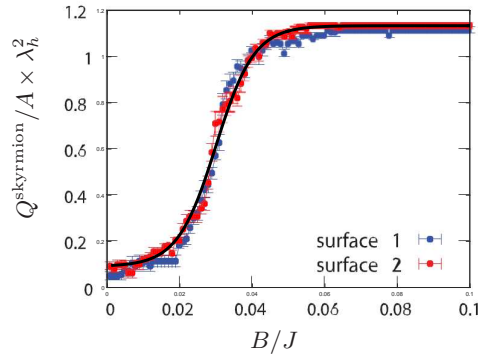}
  \put(-100,-9){$B/J$}
  \put(-190,40){\rotatebox{90}{{$Q^{\text{skyrmion}} /A\times \lambda_{h}^2$}}}
  \caption{(Color online) The evolution curves of the skyrmion numbers on the lower and upper surfaces with respect to the external magnetic field B. The black curve is from our analytical expressions in Eq. \eqref{surface-skyrmion-number}. The dots with error bars are the numerical simulation results presented in Ref.[\onlinecite{2013SCIENCE}].
  }
  \label{Qs_Compare}
\end{figure}

It should be emphasized that the above discussion is totally based on topological argument and statistical analysis, no dynamical details are involved. Therefore, the exact values of the parameters in our results cannot be specified analytically. Nevertheless, by the use of the data fit technique in numerical analysis, we can always compare our analytical expression with the numerical simulation to check the validity of our result. The comparison of numerical simulation  \cite{2013SCIENCE} for the surface skyrmion number and our modified logistic curve in Eq. (\ref{surface-skyrmion-number}) is shown in Fig. \ref{Qs_Compare}, the corresponding parameters $K' =  1.043$, $Q'_0                      =     0.003575$, $\tilde rh =  189.5$ and $Q^s_{\text{\tiny{correction}}}  =   0.09121$ are determined by data fit. An excellent agreement is found, and the goodness of the fit is $R^2= 99.94\%$.

\begin{figure}[h!]
  \centering
  \includegraphics[width=0.70 \linewidth]{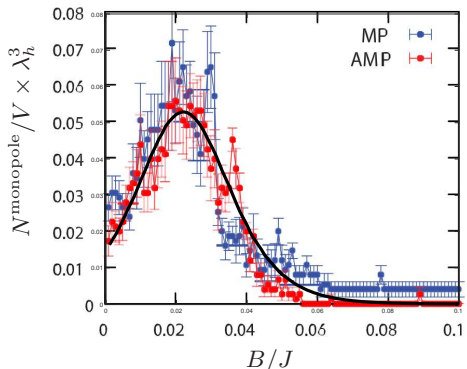}
  \begin{picture}(0,0)
    \put(-100,-9){$B/J$}
    \put(-190,40){\rotatebox{90}{{$N^{\text{monopole}}/V\times \lambda_{h}^3$}}}
  \end{picture}
  \caption{(Color online)The evolution curves of the MP/AMP charge density in the bulk of the chiral magnet with respect to the external magnetic field B. The black curve is from our analytical expressions in Eq. \eqref{monopole_density}. The dots with error bars are the numerical simulation results presented in Ref.[\onlinecite{2013SCIENCE}].
  }
  \label{Qm_Compare}
\end{figure}

From Eq. \eqref{Q monopole 2} we see that a direct calculation of the monopole charge require the information of $K$ and $Q_0$, these can be approximately obtained from the expression of skyrmion number on surface, i.e.
\begin{align}
K &= \lim_{B\rightarrow\infty} Q^S(B) = \lim_{B\rightarrow\infty} Q^S_{\rm boundary}(B) \nonumber\\
Q_0 &= \lim_{B\rightarrow0} Q^S(B) = \lim_{B\rightarrow0} Q^S_{\rm boundary}(B) \nonumber
\end{align}
Thus, we get $K = 1.1336$, $Q_0 = 0.00947$ and $B_0 = \frac{1}{rh} \text{ln}(\frac{K}{Q_0}-1) = 2.3948 \frac{1}{rh}$ for Eq. \eqref{Q monopole 2}.
By taking into account that the simulation results \cite{2013SCIENCE} gives the monopole number density in the bulk, there should be an addition constant factor in Eq. \eqref{Q monopole 2} that reflects the size of the bulk material, hence we have for the monopole number density
\begin{equation}
  n^{MP/AMP} (B) =   C \text{sech}^2 (\frac{rh}{2} B- 1.1974).
  \label{monopole_density}
\end{equation}
The comparison with the numerical result \cite{2013SCIENCE} is exhibited in Fig. \ref{Qm_Compare} with $C = 0.053$ and $rh = 113.5$, the goodness of the fit is $R^2= 95.23\%$. Again, this comparison shows that our analytical result matches well with the numerical simulation.

In conclusion, by the use of the topological current theory, the relationship between the change of skyrmion number and the emergent monopole charge has been presented. It is found that the change of skyrmion number of a layer in the system is equal to the net monopole charges that flows through the layer. Based on this relation, with the help of fundamental statistical argument, the analytical expression of the evolution functions of skyrmions and monopoles in a 3D chiral magnet  with respect to the external magnetic fields have been derived. It is found that the evolution of the skyrmion number in the bulk of the system satisfies the logistic function, while the growth curve of the surface skyrmion number turns out to be a modified logistic curve. Comparison shows that our analytical result is in an excellent agreement with numerical simulations \cite{2013SCIENCE}. Our analytical result may shed a light to the research on skyrmions in chiral magnets as well as topological phase transitions in those systems.

This work was supported by National Natural Science Foundation of China under Grant No. 11275119 and by Ph.D. Programs Foundation of Ministry of Education of China under Grant No. 20123108110004.


\begin{thebibliography}{50}

\bibitem{2013Writing and Deleting Skyrmions} N. Romming {\it et al.}, Science {\bf 341}, 636(2013), and the references therein

\bibitem{2015_skyrmion_logic_gates} X. Zhang, M. Ezawa, and Y. Zhou, Sci. Rep. {\bf 5}, 9400 (2015)

\bibitem{2015Magnetic bubbles with a twist} K. von Bergmann, Science {\bf 349}, 234(2015)

\bibitem{2006skyrmion ground states} U. K. R\"{o}βler,  N. Bogdanov and C. Pfleiderer,
        Nature {\bf 442}, 797-801 (2006).

\bibitem{2014Topological_Changes}R.G. El\'{\i}as and Alberto D. Verga, Phys. Rev. B {\bf 89}, 134405 (2014).

\bibitem{2015Capturing skyrmion}J. M\"{u}ller and A. Rosch, Phys. Rev. B \textbf{91}, 054410 (2015).

\bibitem{2015Blowing skyrmion bubbles} W. Jiang {\it et al.}, SCIENCE {\bf 349}, 283 (2015).

\bibitem{2014dynamics of driven skyrmion}C. Sch\"{u}tte {\it et al.}, Phy. Rev. B {\bf 90}, 174434 (2014)

\bibitem{2015 switching skyrmion}Y. Liu {\it et al.}, Phy. Rev. B {\bf 91}, 094425 (2015)

\bibitem{2015 skyrmion size and shape} N. Romming {\it et al.}, Phys. Rev. Lett. {\bf 114}, 177203 (2015)

\bibitem{2015skyrmion dynamics} S. Komineas and N. Papanicolaou, Phys. Rev. B {\bf 92}, 064412 (2015).

\bibitem{2013NNano}  N. Nagaosa and  Y. Tokura,
        Nat. Nanotechnol. {\bf 8}, 899-911 (2013).

\bibitem{1983_Wilczek_Zee} F. Wilczek and A. Zee, Phys. Rev. Lett. {\bf 51}, 2250 (1983)

\bibitem{1988_Dzyaloshinskii_Polyakov_Wiegmann} I.E. Dzyaloshinskii, A.M. Polyakov, and P.B. Wiegmann, Phys. Lett. A{\bf 127}, 112 (1988); P.B. Wiegmann, Phys. Rev. Lett. {\bf 60}, 821 (1988); A.M. Polyakov, Mod. Phys. Lett. A {\bf 3}, 325 (1988)

\bibitem{1991Dynamics of Magnetic Vortices} N.Papanicolaou and T.N. Tomaras, Nucl. Phys. \textbf{B360} 425-462(1991)

\bibitem{2014_taylor_skyrmion_2D_Nat_Com} B. Dup\'e {\it et al.}, Nat. Comm. {\bf 5}, 4030 (2014)

\bibitem{2010Lorentz microscopy}  X. Z. Yu {\it et al.},
        Nature {\bf 465}, 901-904 (2010).

\bibitem{2011spin-resolved STM} S. Heinze {\it et al.},
        Nature Phys. {\bf 7}, 713-718 (2011)

\bibitem{2015_observation_skyrmion_lattice}T. Tanigaki {\it et al.}, Nano Lett. {\bf 15}, 5438 (2015)

\bibitem{2014_skyrmion_crystal_stiffness} Y. Nii {\it et al.}, Phys. Rev. Lett. {\bf 113}, 267203 (2014)

\bibitem{2009skyrmion lattice} S. M\"{u}hlbauer {\it et al.},
        Science {\bf 323}, 915-919 (2009).

\bibitem{1988_Haldane} F.D.M. Haldane, Phys. Rev. Lett. {\bf 61}, 1029 (1988)

\bibitem{2009YJ} Y. Jiang and G.-H. Yang, Phys. Lett. A {\bf 373}, 4194 (2009).

\bibitem{2013SCIENCE} P. Milde {\it et al.}, SCIENCE \textbf{340}, 1076 (2013).

\bibitem{2016Nature Comm} N. Kanazawa {\it et al.} Nat. Comm. {\bf 7}, 11622 (2016).

\bibitem{2004YJ} Y. Jiang, Phys. Rev. B {\bf 70}, 012501. (2004).

\bibitem{2000YJ} Y. Jiang and Y.S. Duan, J. Math. Phys. {\bf 41}, 2616 (2000); {\it ibid.}, 6463 (2000).

\bibitem{Hatcher_topology} A. Hatcher, {\it Algebraic Topology} (Cambridge University Press, Cambridge, UK, 2002)

\bibitem{2014Dynamics of Emergent Magnetic Monopole} C. Sch\"utte and A. Rosch, Phys. Rev. B {\bf 90}, 174432 (2014).



\end{thebibliography}
\end{document}